\documentclass[12pt]{article}
\usepackage{latexsym,amsmath,amssymb}

\usepackage[sort&compress]{natbib}
\begin{document}

\title{Quantum gravitational contributions to quantum electrodynamics}

\author{David J. Toms\\
School of Mathematics and Statistics,\\
Newcastle University,\\
Newcastle upon Tyne, U.K. NE1 7RU}

\bigskip
\date{To be published in {\it Nature}. Received by journal June 5, 2010.}

\maketitle\eject

{\bf Quantum electrodynamics describes the interactions of electrons and photons. Electric charge (the gauge coupling constant) is energy dependent, and there is a previous claim that charge is affected by gravity (described by general relativity) with the implication that the charge is reduced at high energies. But that claim has been very controversial with the situation inconclusive. Here I report an analysis (free from earlier controversies) demonstrating that that quantum gravity corrections to quantum electrodynamics have a quadratic energy dependence that result in the reduction of the electric charge at high energies, a result known as asymptotic freedom.}     

\bigskip\hrule\bigskip

The standard model of particle physics is based in part on quantised Yang-Mills 
(or non-Abelian) gauge fields.  The quantisation of such fields leads to the important prediction that at high-energy the strength of the gauge coupling constant governing the interaction of the Yang-Mills fields weakens, a phenomenon known as asymptotic freedom~\cite{GrossandWilczek,Politzer}.  This contrasts with a theory like quantum electrodynamics (that describes electrons and photons) where the gauge coupling constant, the electric charge in this case, gets stronger as the energy scale increases.  A practical consequence of this is that for Yang-Mills theories that are asymptotically free a perturbative approach based on a weak coupling constant becomes more reliable at high energy, whereas for quantum electrodynamics perturbation theory breaks down.

The key equations that govern the behaviour of the coupling constants in quantum field theory are the renormalisation group Callan-Symanzik equations~\cite{Callan,Symanzik}. If we let $g$ denote a generic coupling constant, then the value of $g$ at energy scale $E$, the running coupling constant  $g(E)$, is determined by
\begin{equation}\label{CS}
E\frac{dg(E)}{dE}=\beta(E,g),
\end{equation}
where $\beta(E,g)$ is the renormalisation group $\beta$-function. Asymptotic freedom is signalled by $g(E)\rightarrow0$ as $E\rightarrow\infty$, requiring $\beta<0$ in this limit. 

In the standard model of particle physics gravity is usually ignored as it plays an inessential role in most calculations of interest.  Additionally, if we view Einstein's theory of gravity as a fundamental theory it exhibits the undesirable property of non-renormalisability~\cite{tHooftVeltman,DeservanN1,DeservanN2,DeservanN3,DeserTsaovanN}, and hence lacks the power of predictability.  However, it is possible instead to view Einstein gravity as an effective theory that is only valid below some high energy scale~\cite{Donoghue1,Donoghue2}. The cut-off scale is usually associated with the Planck scale $E_P\sim10^{19}$GeV, the natural quantum scale for gravity. Above this energy scale some theory of gravity other than Einstein's theory applies (string theory for example); below this energy scale we can deal with quantised Einstein gravity and obtain reliable predictions~\cite{Donoghue1,Donoghue2,Burgess}. Adopting this effective field theory viewpoint it is perfectly reasonable to include Einstein gravity with the standard model of particle physics and to examine its possible consequences using quantum field theory methods.

With this in mind,  a calculation was performed~\cite{RW} in Einstein-Yang-Mills theory that looked at the effects of quantum gravity on the running Yang-Mills gauge coupling constant.  This calculation found that quantum gravity leads to a correction to the renormalisation group $\beta$-function (not present in the absence of gravity) that tends to result in asymptotic freedom; this holds even for theories (like quantum electrodynamics) that are not asymptotically free when gravity is neglected.  Possible phenomenological consequences of this result have been considered~\cite{Gogoladze}.

The potential importance of the original calculation~\cite{RW} stimulated a number of further investigations that cast doubt on its findings.  It was shown~\cite{Piet} that a different choice of gauge condition led to the absence of any quantum gravity correction to the Yang-Mills $\beta$-function.  Because of the possible gauge condition dependence, a gauge-condition independent version of the background-field method~\cite{Vilkovisky,DeWittVilk} along with dimensional regularisation~\cite{tHooftVeltmandimreg} was used~\cite{Toms1} and also found no quantum gravity contribution to the gauge coupling constant.  The original result~\cite{RW} came from quadratic divergences that automatically get regulated to zero using dimensional regularisation~\cite{tHooftVeltmandimreg}.  

The situation was analysed with a traditional Feynman diagram approach using standard Feynman rules and it was shown~\cite{Ebertetal} that if a momentum space cut-off was used the quadratic divergences could be made to cancel leaving a result that was consistent with dimensional regularisation.  On the other hand a novel method of regularisation, called loop regularisation, claimed~\cite{TangWu} that the quadratic divergences were present, contradicting earlier results~\cite{Piet,Toms1,Ebertetal}.  Subsequent work has looked at the role of the cosmological constant~\cite{Toms2,Toms3}, corrections in higher dimensions~\cite{EbertetalJHEP}, and scalar~\cite{RodigastSchuster2,MackayToms} and Yukawa fields~\cite{RodigastSchuster2}.  Additionally, related calculations have been performed in the exact renormalisation group approach~\cite{RG1,RG2,RG3}.  

The issue of quadratic divergences and the possible gauge condition dependence is the central theme the present paper. The generalised background-field method~\cite{Vilkovisky,DeWittVilk} that is used here differs from the usual one by the addition of an extra contribution that is essential for maintaining gauge condition invariance. The difference between the standard and generalised effective actions can be controlled by a single parameter $v$ that when set equal to one gives the gauge condition independent result, and when set to zero gives the standard result that coincides with a traditional Feynman diagram analysis. Gauge invariance is maintained by using a proper time~\cite{Schwinger,dynamical} cut-off. The importance of this choice of regularisation is that it is capable of revealing any possible quadratic divergences unlike many other methods. It will be shown that the quadratic divergences that are obtained do not depend on the parameter $v$, and that the traditional background-field method leads to a gauge condition dependent result for the charge renormalisation.  The contribution of the extra terms needed to obtain the gauge condition independent result is greatly simplified~\cite{FradkinTseytlin} by choosing a particular gauge condition described below. Because the generalised effective action is independent of the choice made for the gauge condition there is no generality lost by this procedure.  (A comprehensive review and justification of the method can be found in a recent monograph~\cite{ParkerToms}.) Surprisingly, it will be shown that the quadratic divergences do not cancel, and that the basic prediction originally found~\cite{RW} is correct, although with a slight modification of the $\beta$-function. Quantum electrodynamics is found to be asymptotically free if it is coupled to quantum gravity.

\section*{Effective action}
The model studied is Einstein gravity with a cosmological constant coupled to quantum electrodynamics in four spacetime dimensions. (The presence of the cosmological constant is not essential to the point that will be made, but is included for generality.) The basic fields are the spacetime metric $g_{\mu\nu}$, the electromagnetic field $A_{\mu}$, and the Dirac field $\psi$. The principle aim is to calculate the quantum gravity contribution to the renormalisation of the charge. To do this it is sufficient to adopt the background-field method, choose the background metric to be flat, the background Dirac field to vanish, the background electromagnetic field to correspond to a constant field strength $\bar{F}_{\mu\nu}$, and to concentrate on that part of the one-loop effective action that is divergent and involves the square of the background field strength. 

A standard calculation shows that the effective action to one-loop order is given by
\begin{equation}\label{oneloop}
\Gamma^{(1)}=\frac{1}{2}\ln\det\Delta^{i}{}_{j}-\ln\det Q_{\alpha \beta}
-\ln\det(i\gamma^{\mu}\partial_\mu+e\gamma^\mu\bar{A}_{\mu}-im).
\end{equation}
The last term (with $\bar{A}_{\mu}$ the background gauge field) is the result of performing a functional integral over the Dirac field. The middle term is the contribution from the ghost fields required to remove the unphysical degrees of freedom of the gravity and electromagnetic fields. The first term is the result of integrating over the spacetime metric and electromagnetic fields; $\Delta^{i}{}_{j}$ is a second order differential operator that can be found from earlier work~\cite{Toms2,Toms3} and will not be written down here due to its complexity. It is found by expanding the classical Einstein-Maxwell action about the background fields using $g_{\mu\nu}=\delta_{\mu\nu}+\kappa h_{\mu\nu}$ and $A_{\mu}=\bar{A}_\mu+a_{\mu}$. A Riemannian metric and standard conventions~\cite{TomsYM} for the Dirac fields are chosen. $\kappa^2=32\pi G$ with $G$ Newton's constant. 

The gauge-condition independent background-field method~\cite{Vilkovisky,DeWittVilk} is used as described in the final paragraph of the introduction. The gauge conditions adopted are (with $h=\delta^{\mu\nu}h_{\mu\nu}$)
\begin{eqnarray}
\chi_\lambda&=&\partial^\nu h_{\lambda\nu}-\frac{1}{2}\partial_{\lambda}h+\frac{\kappa}{2}\omega\left(\bar{A}_\lambda\partial^\mu a_\mu+\bar{F}_{\mu\lambda}a^\mu\right),\label{gaugegravity}\\
\chi&=&\partial^\mu a_\mu.\label{gaugeelectro}
\end{eqnarray}
To obtain the gauge condition independent result it is essential to take $\omega=1$ in our calculation. However $\omega$ is kept general at this stage to indicate the gauge condition dependence of standard methods of calculation. The choice $\omega=0$ is often made and is called the de Donder gauge. 

The gauge conditions are incorporated in the usual way by altering the action with a gauge fixing term
\begin{equation}
S_{GF}=\frac{1}{\xi}\int d^4x\;\chi^{\lambda}\chi_{\lambda}+\frac{1}{2\zeta}\int d^4x\;\chi^2.
\end{equation}
Here $\xi$ and $\zeta$ are two dimensionless parameters. The choice $\xi=1=\zeta$ is often made because it simplifies the calculation enormously. However in doing this you lose any hope of addressing the gauge condition dependence of the calculation. $\xi$ and $\zeta$ will be kept general here noting that the gauge condition independent effective action is the same as what is found by taking the limits $\xi\rightarrow0$ and $\zeta\rightarrow0$ (along with $\omega=v=1$).

\section*{Heat kernel}
For operator $\Delta^{i}{}_{j}$ the heat kernel $K^{i}{}_{j}(x,x';\tau)$ is defined by
\begin{equation}\label{heatK}
-\frac{\partial}{\partial\tau}K^{i}{}_{j}(x,x';\tau)=\Delta^{i}{}_{k}K^{k}{}_{j}(x,x';\tau),
\end{equation}
with boundary condition $K^{i}{}_{j}(x,x';\tau=0)=\delta^{i}_{j}\delta(x,x')$. $\tau$ is called the proper time~\cite{Schwinger,dynamical}. The Green function $G^{i}{}_{j}(x,x')$ for the operator $\Delta^{i}{}_{j}$ is
\begin{equation}\label{Green}
\Delta^{i}{}_{k}G^{k}{}_{j}(x,x')=\delta^{i}_{j}\delta(x,x').
\end{equation}
It follows that the Green function and heat kernel are related by
\begin{equation}\label{GinK}
G^{i}{}_{j}(x,x')=\int\limits_{0}^{\infty}d\tau\,K^{i}{}_{j}(x,x';\tau).
\end{equation}

The importance of the heat kernel for quantum field theory arises from the existence~\cite{dynamical,Gilkey} of an asymptotic expansion as $\tau\rightarrow0$:
\begin{equation}\label{asymptotic}
K^{i}{}_{j}(x,x;\tau)\sim(4\pi\tau)^{-n/2}\sum_{r=0}^{\infty}\tau^rE_{r}{}^{i}{}_{j}(x)
\end{equation}
where $n$ is the spacetime dimension (chosen as 4 here) and the heat kernel coefficients $E_{r}{}^{i}{}_{j}(x)$ depend only locally on the details of coefficients of the differential operator $\Delta^{i}{}_{j}$. For many operators of interest in physics the results of the first few coefficients are known~\cite{dynamical,Gilkey}; however, the operators needed for the present calculation are more general than considered so far. (Some checks on our results follow from a different method~\cite{BV}.)

The divergent part of the effective action, as well as the Green function, can be related to the heat kernel coefficients. Formally (before regularisation)
\begin{equation}\label{G1}
L_{\Delta}=\frac{1}{2}\ln\det\Delta^{i}{}_{j}=-\frac{1}{2}\int \,d^nx\int\limits_{0}^{\infty}\frac{d\tau}{\tau}{\rm tr} K^{i}{}_{j}(x,x;\tau).
\end{equation}
The one-loop effective action (\ref{oneloop}) is then given by
\begin{equation}\label{gammaoperator}
\Gamma^{(1)}=L_{\Delta}-2L_Q-2L_{\rm Dirac}.
\end{equation}
As with the Green function (\ref{Green}) the divergent part of (\ref{G1}) comes from the $\tau\sim0$ limit of the proper time integral. To deal with this in a way that respects gauge invariance, general coordinate invariance, and the structure of any quadratic divergences, a proper time cut-off $\tau_c$ is used where the lower limit of the proper time integration is replaced with $\tau_c$. The divergences will show up~\cite{Schwinger} as $\tau_c\rightarrow0$. Because of this the divergent part ${\rm divp}(L_\Delta)$ of $L_\Delta$ follows from using the asymptotic form of the heat kernel expansion (\ref{asymptotic}). 

In order to make contact with the standard renormalisation group procedure that uses an energy scale cut-off note that $\tau_c\sim({\rm length})^2=({\rm energy})^{-2}$ in $\hbar=1=c$ units. The proper time cut-off can therefore be replaced with an energy cut-off $E_c$ using $\tau_c=E_c^{-2}$. The divergent part of $L_\Delta$ becomes
\begin{equation}\label{divpGamma}
{\rm divp}(L_\Delta)=-\frac{1}{32\pi^2}\int d^4x\Big\lbrace \frac{1}{2}E_c^{4} {\rm tr}E_0+E_c^{2}{\rm tr}E_1
+{\rm tr}E_2\,\ln E_c^2\Big\rbrace.
\end{equation}
Previous work~\cite{Toms1,Toms2,Toms3,RodigastSchuster2,MackayToms} used dimensional regularisation~\cite{tHooftVeltmandimreg} with the spacetime dimension taken as $n=4+\epsilon$ with $\epsilon\rightarrow0$ understood. In this case the lower limit on the proper time integration can be kept as $\tau=0$ and it is found that the divergent part of the effective action $L_\Delta$ contains a simple pole as $\epsilon\rightarrow0$ given by
\begin{equation}\label{dimreg}
{\rm divp}(L_\Delta)=\frac{1}{16\pi^2\epsilon}\int d^4x\,{\rm tr}E_2.
\end{equation}

Comparison of the dimensional regularisation result (\ref{dimreg}) with that of the cut-off method (\ref{divpGamma}) shows that the coefficient of the simple pole at $\epsilon=0$ in dimensional regularisation is the same as the coefficient of $\ln E_c^{-1}$. The quartic divergence (proportional to $E_c^4$), and the quadratic divergence (proportional to $E_c^2$) do not appear in dimensional regularisation; they are both regulated to zero. If the cut-off energy $E_c$ is regarded as a fundamental scale in the effective field theory, then neglect of these terms could be significant. For the consideration of the gauge coupling constant renormalisation it is possible to show by calculating the $E_0$ coefficient that there can be no contribution to the charge renormalisation from the term involving $E_c^4$ in (\ref{divpGamma}). However there is a potential contribution from the quadratic divergence (middle term of (\ref{divpGamma})); in fact, it will be demonstrated that such a divergence is present, as found in the original calculation~\cite{RW}, and that it alters the result found by using dimensional regularisation.

It is clear from (\ref{divpGamma}) that the central object of importance for deciding whether or not quadratic divergences are present lies in the expression ${\rm tr}E_1$. Although ${\rm tr}E_1$ is known for special operators~\cite{dynamical,Gilkey,BV} it is not known for the general operators that arise in the present calculation. The full details are technically involved and will be presented elsewhere. If dimensional regularisation is chosen, with $n=4+\epsilon$, then $G^{i}{}_{j}(x,x)$ has a simple pole~\cite{Tomsscalar} as $\epsilon\rightarrow0$ whose residue involves the heat kernel coefficient $E_1$:
\begin{equation}\label{poleGreen}
{\rm divp}\left(G^{i}{}_{j}(x,x)\right)=-\frac{1}{8\pi^2\epsilon}\;E_{1}{}^{i}{}_{j}(x).
\end{equation}
It should be noted that dimensional regularisation is only used as a technical device for calculating the heat kernel coefficient $E_1$, and that this is distinct from any choice of regularisation employed for the effective action. The calculation of $E_1$ has been checked using another method that does not use dimensional regularisation.

The heart of the calculation now involves finding the pole part of $G^{i}{}_{j}(x,x)$ and then reading off the expression for $E_1$. To accomplish this the local momentum space approach~\cite{BunchParker} is adopted that utilises a normal coordinate expansion of the operator $\Delta^{i}{}_{j}$. The general form of $\Delta^{i}{}_{j}$ is
\begin{equation}\label{Delta}
\Delta^{i}{}_{j}=(A^{\alpha\beta})^{i}{}_{j}\partial_\alpha\partial_\beta+ (B^{\alpha})^{i}{}_{j}\partial_\alpha+(C)^{i}{}_{j}
\end{equation}
for coefficients $(A^{\alpha\beta})^{i}{}_{j},(B^{\alpha})^{i}{}_{j}$ and $(C)^{i}{}_{j}$ that depend on the spacetime coordinates through the background field. Normal coordinates are introduced at $x'$ with $x^\mu=x^{\prime\mu}+y^\mu$ and all of the coefficients in (\ref{Delta}) are expanded about $y^\mu=0$. This gives
\begin{equation}\label{Aexp}
(A^{\alpha\beta})^{i}{}_{j}=(A_{0}^{\alpha\beta})^{i}{}_{j} +\sum_{n=1}^{\infty}(A^{\alpha\beta}{}_{\mu_1\cdots\mu_n})^{i}{}_{j}y^{\mu_1}\cdots y^{\mu_n}
\end{equation}
with similar expansions for $(B^{\alpha})^{i}{}_{j}$ and $(C)^{i}{}_{j}$. The Green function is Fourier expanded as usual,
\begin{equation}\label{Greenexp}
G^{i}{}_{j}(x,x')=\int\frac{d^np}{(2\pi)^n}e^{ip\cdot y}G^{i}{}_{j}(p),
\end{equation}
except that the Fourier coefficient $G^{i}{}_{j}(p)$ can also have a dependence on the origin of the coordinate system $x'$ that is not indicated explicitly. 

By substituting (\ref{Greenexp}) into (\ref{Green}) and using the expansion (\ref{Aexp}) and similar ones for $B^\alpha$ and $C$ it is possible to solve for $G^{i}{}_{j}(p)$ as an asymptotic series in $1/p$ beginning at order $p^{-2}$. If
\begin{equation}\label{Gpexp}
G^{i}{}_{j}(p)=G_{0}{}^{i}{}_{j}(p)+ G_{1}{}^{i}{}_{j}(p) + G_{2}{}^{i}{}_{j}(p) +\cdots
\end{equation}
where $G_{r}{}^{i}{}_{j}(p)$ is of order $p^{-2-r}$ as $p\rightarrow\infty$ it is easy to see that to calculate the pole part of $G^{i}{}_{j}(x,x)$ as $n\rightarrow4$ only terms up to and including $G_{2}{}^{i}{}_{j}(p)$ are needed. These can be found iteratively beginning with $G_{0}{}^{i}{}_{j}(p)$. The evaluation of these terms is extremely complicated and most of the brute force calculation was done using Cadabra~\cite{cadabra}. 

The net result of this lengthy calculation is that the gravity and gauge field contributions result in
\begin{eqnarray}
{\rm tr}E_1&=&\kappa^2\Big(\frac{3}{8}-\frac{3}{4}\omega+\frac{1}{8}\omega^2+\frac{3}{8}\xi -\frac{1}{2}\omega\xi+\frac{3}{8}\omega\zeta
-\frac{1}{32}\omega\xi\zeta+\frac{1}{32}\omega^2\zeta\Big)\bar{F}^2\nonumber\\
&&+(12+8\xi^2+3v+v\zeta^2)\Lambda.\label{E1G}
\end{eqnarray}
One check on this result is that all terms that involve $1/\xi$ that arise at intermediate stages of the calculation cancel from the final result. Another useful check is that if the choice $\xi=1=\zeta$ is made then the result for $E_1$ follows from a standard result~\cite{Gilkey} and agrees with the result for ${\rm tr} E_1$ found above. It is noteworthy that the parameter $v$ that marks the difference between the standard and gauge condition independent effective actions does not enter the term in the  $E_1$ coefficient that multiplies the field strength $\bar{F}^2$, and therefore cannot contribute to the quadratic divergences responsible for charge renormalisation. This was noted earlier~\cite{Toms3} using a different approach. (The parameter $v$ does occur in the term involving the cosmological constant, but this can play no role in charge renormalisation.) To get the result of the standard background-field method, that must also agree with an analysis using Feynman diagrams and normal Feynman rules, we simply set $v=0$. The coefficient of $\bar{F}^2$ in this term can be seen to depend explicitly on the two gauge parameters $\xi$ and $\zeta$, as well as on the gauge condition parameter $\omega$. In this gauge condition dependent case it still follows that there is a quadratic divergence, a result that is at variance with using a momentum space cut-off~\cite{Ebertetal}.

A similar procedure can be applied to the ghost field contribution (second term of (\ref{oneloop})) and the Dirac field contribution (last term of (\ref{oneloop})). The Dirac field has no quadratic divergence since~\cite{TomsYM} ${\rm tr}E_1=0$, but does have a logarithmic divergence coming from  ${\rm tr}E_2$. The ghost field operator results in \begin{equation}\label{ghost}
{\rm tr}E_1=-\frac{\kappa^2}{4}\omega^2\bar{F}^2.
\end{equation}

\section*{Divergences and renormalisation group}
The overall result for the quadratically divergent part of the complete one-loop effective action (\ref{gammaoperator}) that involves $\bar{F}^2$ is
\begin{equation}\label{complete}
\Gamma_{\rm quad}^{(1)}=-\frac{\kappa^2 E_c^2}{32\pi^2}\Big( \frac{3}{8}-\frac{3}{4}\omega+\frac{5}{8}\omega^2+\frac{3}{8}\xi -\frac{1}{2}\omega\xi +\frac{3}{8}\omega\zeta
-\frac{1}{32}\omega\xi\zeta+\frac{1}{32}\omega^2\zeta\Big)\int d^4x\bar{F}^2.
\end{equation}
This conclusively demonstrates the gauge parameter and gauge condition dependence of the standard result. If $\xi\rightarrow0,\zeta\rightarrow0,\omega\rightarrow1$ are taken to obtain the gauge condition independent result as discussed above, the non-zero result
\begin{equation}\label{correct}
\Gamma_{\rm quad}^{(1)}=-\frac{\kappa^2 E_c^2}{128\pi^2}\int d^4x\bar{F}^2
\end{equation}
is found.

This is not the complete divergent part of the effective action that involves $\bar{F}^2$ because there is still the logarithmic contribution. It is possible to extend the method of calculation described above to calculate $E_2$, but the results can be deduced from the known~\cite{Toms2,Toms3} poles found for the first two terms of (\ref{gammaoperator}), and for the Dirac contribution~\cite{TomsYM}. The net result for the divergent part of the effective action that involves $\bar{F}^2$ and therefore contributes to charge renormalisation is
\begin{equation}
{\rm divp}(\Gamma^{(1)})=\left(-\frac{\kappa^2E_c^2}{128\pi^2} -\frac{3\kappa^2\Lambda}{256\pi^2}\ln E_c^2 +\frac{e^2}{48\pi^2}\ln E_c^2\right)\int d^4x\bar{F}^2.\label{gammadiv}
\end{equation}
From this the renormalisation group function in (\ref{CS}) that governs the running electric charge to be calculated to be
\begin{equation}\label{betafcn}
\beta(E,e)=\frac{e^3}{12\pi^2}-\frac{\kappa^2}{32\pi^2}(E^2+\frac{3}{2}\Lambda)e.
\end{equation}
The first term on the right hand side of (\ref{betafcn}) is that present in the absence of gravity (found by letting $\kappa\rightarrow0$) and results in the electric charge increasing with the energy. The second term on the right hand side of (\ref{betafcn}) represents the correction due to quantum gravity. For pure gravity with no cosmological constant, or for small $\Lambda$ as present observational evidence suggests~\cite{Wmap}, the quantum gravity contribution to the renormalisation group $\beta$-function is negative and therefore tends to result in asymptotic freedom, in agreement with the original calculation~\cite{RW}.

\section*{Outlook}

Although the calculation has been done for quantum electrodynamics, similar conclusions follow for Yang-Mills fields with and without further matter present. It is possible that the running of the gauge coupling constants (found by solving (\ref{CS}) for a realistic gauge theory) can lower the unification scale in comparison to what is expected in the absence of gravity~\cite{RW}. In addition, if we allow spacetime to have more than four dimensions, then the presence of extra dimensions could lower the effective gravitational scale below the Planck scale~\cite{ADD} and render the results of running charges measurable~\cite{Gogoladze} in the LHC. There is also an intriguing connection with the weak gravity conjecture~\cite{weakgravity} which predicts that the natural cut-off should be a couple orders of magnitude below the Planck scale. The weak gravity conjecture translates into the requirement that the gravitational contribution to the renormalisation group $\beta$-function should be less than that found in the absence of gravity~\cite{Huang}.

\bigskip
\noindent{\bf Author Information} Reprints and permissions information is available at npg.nature.com/reprintsandpermissions. Correspondence should be addressed to the author (d.j.toms@newcastle.ac.uk).

\end{document}